\documentstyle[times,epsfig]{jaa}

\begin{document}

\title[The hard X-ray emission of GX~17+2]{The hard X-ray emission along the ``Z'' track in GX~17+2}
%\title[]{The hard X-ray emission along the ``$\nu$'' track in GX~17+2}

\author[G. Q. Ding \& C. P. Huang]{G. Q. Ding$^{1,*}$ \& C. P. Huang$^{1}$ \\
%\author[]{Chunping Huang$^{1}$ \& Guoqaing Ding$^{1,*}$\\
        $^1$Xinjiang Astronomical Observatory, CAS, 150 Science 1-Street, Urumqi 830011, China \\
        $^*$e-mail: dinggq@xao.ac.cn, dinggq@gmail.com\\}

%\author[]{Chunping Huang$^{1}$ \& Guoaing Ding$^{1,}$\thanks{e-mail:dinggq@xao.ac.cn, dinggq@gmail.com} \\
%         $^1$Xinjiang Astronomical Observatory, CAS, 150 Science 1-Street, Urumqi 830011, China}

%\pubyear{1995}
%\volume{16}
%\setcounter{page}{217}
%\date{Received 1996 June 20; accepted 2000 March 09}

\maketitle

\begin{abstract}Using the data from the Proportional Counter Array (PCA) and the 
High-Energy X-ray Timing Experiment (HEXTE) on board {\it Rossi X-Ray Timing Explorer} 
for Z source GX 17+2, we investigate the evolution of its PCA spectra and HEXTE spectra 
along a ``Z'' track on its hardness-intensity diagram. A hard X-ray tail is detected in 
the HEXTE spectra. The detected hard X-ray tails are discontinuously scattered throughout 
the ``Z'' track. The found hard tail hardens from the horizontal branch, through the 
normal branch, to the flaring branch in principle and it contributes $\sim$(20-50)\% 
of the total flux in 20-200 keV. Our joint fitting results of the PCA+HEXTE spectra 
in 3-200 keV show that the portion of Comptonization in the bulk-motion Comptonization 
(BMC) model accounts for the hard X-ray tail, which indicates that the BMC process could 
be responsible for the detected hard tail. The temperature of the seed photons for BMC is 
$\sim$2.7~keV, implying that these seed photons might be emitted from the surface of the 
neutron star (NS) or the boundary layer between the NS and the disk and, therefore, this 
process could take place around the NS or in the boundary layer.         
\end{abstract}


\section{Introduction}

In the past several decades, the huge numbers of data from X-ray satellites make it 
possible to deeply investigate the spectra of X-ray binaries, including black holes 
X-ray binaries (BHXBs) and neutron star X-ray binaries (NSXBs). Usually, the hard 
X-ray tails, i.e. the X-ray spectra higher than dozens of keV, e.g. above $\sim$40 keV, 
have been observed in BHXBs. It is a nature assumption that the hard tails are resulted 
from Comptonization of seed photons by high-energy electrons. In BHXBs, the high-energy 
electrons exist in two scenarios (see review of Done et al. 2007). In the hard 
states of BHXBs, the energetic electrons are assumed to exist in the optically thin 
part between the black hole and an optically thick truncated disk far from the black 
hole, where the high-energy electrons are thermalized and share Maxwellian distribution, 
and the emitting spectrum can be described by thermal Comptonization in which the soft 
photons from the disk are inversely Comptonized by these thermal electrons 
(Shapiro et al. 1976; Sunyaev \& Truemper 1979; Sunyaev \& Titarchuk 1980; 
Titarchuk 1994). While, in the soft states the high-energy electrons are described by 
a ``Hybrid'' distribution, i.e. a mixture with thermal distribution and non-thermal 
power-law (PL) distribution (Poutanen \& Coppi 1998; Coppi 1999; 
Gierli{\'n}ski et al. 1999; Zdziarski et al. 2001); the thermal electrons might come 
from the remains of the hot inner flow and the non-thermal electrons might originate 
from the magnetic flares above the disk (Done et al. 2007). Another alternative 
mechanism for generation of hard X-ray tails in BHXBs is the the bulk-motion 
Comptonization (BMC), in which the seed photons are inversely Comptonized by the 
electrons which freely fall with relativistic speed onto the black 
holes (Titarchuk et al. 1997; Laurent \& Titarchuk 1999). 

NSXBs are divided into two types, i.e. atoll sources and Z sources (Hasinger \& 
van der Klis 1989). The hard X-ray tails have been detected in a few atoll sources. 
The presence of a PL hard tail in the broadband spectra of atoll sources can trace 
back to such detection from the {\it BeppoSAX} observations for 4U 0614+091 
(Piraino et al. 1999). With {\it INTEGRAL} observations, Paizis et al. (2006) 
detected a hard tail in GX 13+1, which makes it the first atoll source with a hard 
tail. Then, this non-thermal PL hard tail with $\Gamma$=2.76 was detected in the 
{\it INTEGRAL} spectrum of 4U 1636-53 during its soft state (Fiocchi et al. 2006). 
A PL hard tail with photon index of $\sim$2.4 was also found in the {\it INTEGRAL} 
spectrum of 4U 1820-30 during its island state (IS) (Tarana et al. 2007). 
Afterward, the similar hard tail having a photon index ($\Gamma$) $\sim$2.9 and 
contributing $\sim$11\% of the total flux in the 0.1-200 keV energy interval was 
detected in 4U 1705-44 from a {\it BeppoSAX} observation (Piraina et al. 2007). 
With the observations of {\it Suzaku} and {\it Rossi X-Ray Timing Explorer} 
({\it RXTE}), such high-energy X-ray tail was detected in Aql X-1 too (Raichur 
et al. 2011), also found in the {\it INTEGRAL} spectrum of another atoll source 
4U 1728-34 (Tarana et al. 2011). Although some models have been proposed to 
explain the detected hard X-ray tails in atoll sources, there has no a unified 
explanation. Paizis et al. (2006) used the generic Comptonization model, i.e. 
the BMC model, to fit the spectrum of GX 13+1 in $\sim$(20-80) keV and got 
statistically good fitting, indicating that the BMC process could be an 
alternative mechanism for producing the hard X-ray emission in this source. 
Due to the high ratio of the radio flux to the peak X-ray flux found 
in 4U 1636-53, Fiocchi et al. (2006) considered that the synchrotron emission 
in the jet could be responsible for the detected hard tail. However, in order 
to explain the high-energy X-ray tails in 4U 1820-30 and 4U 1728-34, 
Tarana et al. (2007, 2011) proposed that the emission of non-thermal electrons 
in a hybrid plasma might be the origin of the detected hard X-ray tails. 

However, the detection of hard X-ray tails in Z sources is unpredictable and the 
detected hard tails present complicated behaviors. Generally, among the Z sources, 
the hard tail is detected on the horizontal branch (HB) and it gradually fades away 
as they moves along the ``Z'' track on their hardness-intensity diagrams (HIDs) 
(Di Salvo et al. 2000, 2002; Lin et al. 2009; Ding et al. 2003, 2011). But, 
occasionally, the hard X-ray emission is not correlated with the positions on the 
HIDs (D'Amico et al. 2001). Analyzing the spectra of High-Energy X-ray Timing 
Experiment (HEXTE) on board {\it RXTE}, D'Amico et al. (2001) found that the 
PL hard X-ray tail of Sco X-1 hardens in the sequence HB$\rightarrow$NB$\rightarrow$FB 
on its HID. Using the {\it RXTE} observations for Sco X-1, D'A{\'i} et al. (2007) 
studied the evolution of broadband spectra (3-200 keV) of Sco X-1 on its color-color 
diagram (CD) and proposed the ``Hybrid'' Comptonization model to explain the hard X-ray 
emission. With {\it INTEGRAL} observation, Revnivtsev et al. (2014) investigated the 
hard tail of Sco X-1. Analyzing the broadband spectra (0.1-200 keV) of a {\it BeppoSAX} 
observation for GX 349+2, Di Salvo et al. (2001) detected a hard tail with photon index 
$\sim$2. A hard X-ray tail of GX 340+0 was also detected in its {\it BeppoSAX} spectra 
(Lavagetto et al. 2004). Using the {\it BeppoSAX} observations for Cir X-1, which hosts 
an elliptical orbit with high eccentricity ($e$ $\sim$ 0.7--0.9), Iaria et al. (2001, 2002) 
studied its broadband spectra near the periastron (orbital phase 0) or at the orbital 
phases close to the apoastron (orbital phase 0.5) and detected similar hard X-ray tails 
with photon index $\sim$3.2 during the two orbital episodes. Interestingly, analyzing 
{\it RXTE} data, Ding et al. (2006b) found orbital modulation for the hard X-ray emission 
of Cir X-1. In this peculiar source, from the periastron to the apastron, the PL hard 
tail softens and its flux increases, and, however, from the apastron to the next 
periastron (orbital phase 1), it hardens and its flux increases. 

The origin of hard tails in Z sources is debated. Some models have been proposed 
for the production of hard X-ray emission in Z sources. Z sources share neutron 
star (NS) surface magnetic field strength of $\sim$$10^9\ {\rm G}$ (Fock 1996; Ding 
et al. 2006b, 2011), so synchrotron emission of energetic electrons might be an 
origin (Riegler 1970; D'Amico et al. 2001; Iaria et al. 2002). Di Salvo et al. 
(2006) proposed the ``Hybrid'' model for the production of hard tails in Z sources. 
In the ``Hybrid'' model, the high-energy electrons are from a hybrid 
thermal/nonthermal corona. A jet might be another origin of hard tails in Z sources 
due to the fact that in these sources both the hard X-ray emission flux and radio 
flux are highest on the HB, then fade away along the ``Z'' track on their HIDs, and, 
finally, become lowest on the flaring branch (FB) (Penninx et al. 1988; Di Salvo 
et al. 2000, 2002; Lin et al. 2009; Ding et al. 2003, 2011; Migliari et al. 2007; 
Fender et al. 2007).   

It is noted that for the past about ten years the BMC model has been considered 
for the production of hard tails in Z sources. Paizis et al. (2006) used this model 
to fit the broadband spectra of NSXBs for the first time, then Farinelli et al. 
(2007, 2008, 2009) developed this model and used it to fit the broadband spectra of 
Z sources. Ding et al. (2011) used the model to fit the broadband spectra of the 
transient Z source XTE J1701-462 and suggested that the bulk-motion Comptonization 
near the NS could be an alternative mechanism for producing the hard tails in this 
source.

In this work, using the data from the Proportional Counter Array (PCA) and the HEXTE 
on board {\it RXTE} satellite for Z source GX 17+2, we study the evolution of its PCA 
spectra and HEXTE spectra along a ``Z'' track on its HID. We detect the hard X-ray 
tails in its  HEXTE spectra and investigate the possible mechanism for producing the 
hard X-ray emission. We describe our data analysis in Section 2, present and discuss 
our results in Section 3, and give our conclusion in Section 4.
      
\section{Data analysis}

\subsection{The selected observations}

\begin{figure*}[t]
\vspace{5.cm}
\centerline{
\includegraphics[width=8cm,height=8cm,angle=-90]{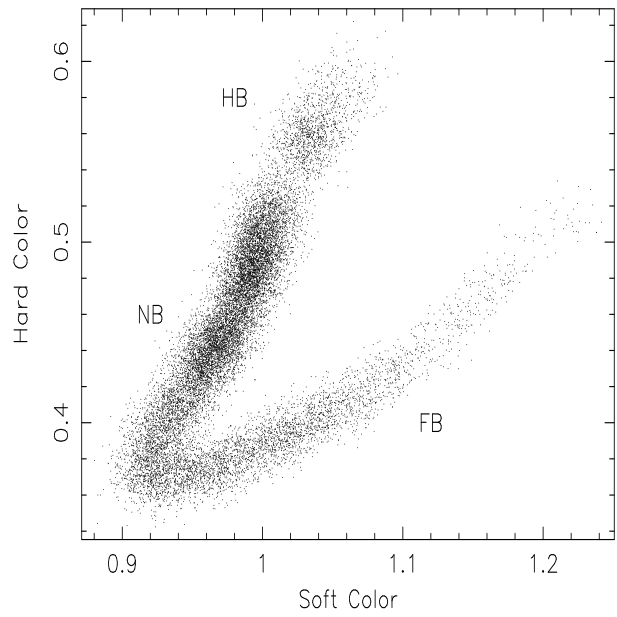}
}
\vspace{0.25cm}
\caption{The CD of GX~17+2. Each point represents 16 s 
background-subtracted data. The soft color is defined as the count rate 
ratio between 4.6--7.1 keV and 2.9--4.6 keV energy bands, while the hard 
color is defined as the count rate ratio between 10.5--19.6 and 7.1--10.5 
energy bands.}\label{fig:CCD}
\end{figure*}

\begin{figure*}[t]
\vspace{5.cm}
\centerline{
\includegraphics[width=8.5cm,height=8cm,angle=0]{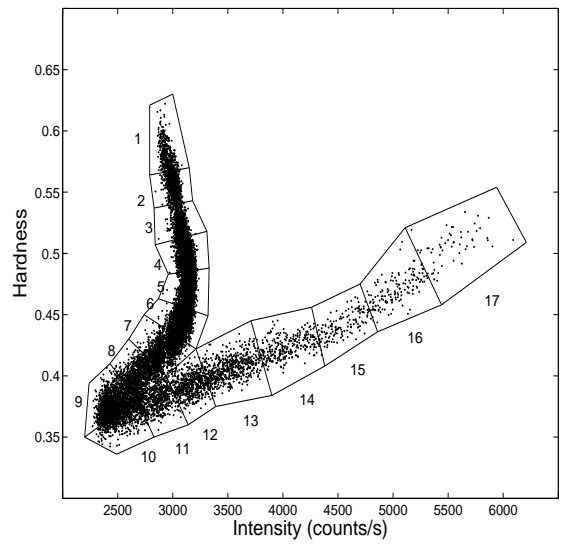}
}
\vspace{0.25cm}
\caption{The HID of GX~17+2, which is divided into seventeen regions as 
labeled. Each point represents 16 s background-subtracted data. The hardness 
is defined as the count rate ratio between 10.5--19.6 keV and 7.1--10.5 keV 
energy bands, and the intensity is defined as the count rate 
in 2.9--19.6 keV.}\label{fig:HID}
\end{figure*}

Using software HEASOFT 6.11 and FTOOLS V6.11, we select the {\it RXTE} 
observations for GX~17+2 between 1999 Oct.~3 and Oct.~12, a total of 
$\sim$297.6~ks, to perform our analysis. The ``Standard 2'' mode data with a 
time resolution of 16 s are used to produce the color-color diagram (CD) and 
the HID. The two proportional counter units (PCUs 0, 2) are used because they 
are always on. Following Homan et al. (2002), the soft color is defined as the 
count rate ratio between the [4.6-7.1 keV (channels 8-13)] and [2.9-4.6 keV 
(channels 4-7)] energy bands, and the hard color as that between the [10.5-19.6 
keV (channels 22-42)] and [7.1-10.5 keV (channels 14-21)] energy bands, and 
intensity as the count rate covering the energy range [2.9-19.6 keV (channels 
4-42)]. Using ``Standard 2'' mode data, applying FTOOLS SAEXTRCT, we extract 
the source light curves. Then, using the background model provided by {\it RXTE} 
team and applying RUNPCABACKEST, a tool of FTOOLS, we produce the background 
files and thus extract the background light curves. Finally, we obtain the 
background-subtracted light curves with various energy bands and create the CD 
and HID by using FTOOLS LCURVE, which are shown by Figure~\ref{fig:CCD} and 
Figure~\ref{fig:HID}, respectively, presenting complete ``Z'' tracks, respectively. 
From the top left to bottom right on the CD, the track is divided into three 
segments which are called HB, normal branch (NB), and FB, respectively. These 
branches are correspondingly demonstrated on the HID (Figure~\ref{fig:HID}). 
We split the ``Z'' track on the HID into 17 regions in order to minimize count 
rate variations in each region and then extract the PCA and HEXTE spectra of 
each of the 17 regions. In Figure~\ref{fig:HID}, the HB, NB, and FB consist of 
regions 1--4, regions 5--9, and regions 10--17, respectively.

\subsection{Extracting the spectra}

Firstly, according to the hard color and intensity of each region in
Figure~\ref{fig:HID}, we determine its absolute time interval. Then, using
SAEXTRCT, we extract the PCA and HEXTE spectra of each region on the ``Z'' 
track by confining the time interval. The ``Standard 2'' mode data from PCUs
0, 2, providing count spectra each 16 s in 129 energy channels covering
2--100 keV, are used to generate the PCA spectra. Using RUNPCABACKEST and
applying the background model provided by {\it RXTE} team, we produce the PCA
background files and, thus, the background spectra and background-subtracted 
spectra of PCA are generated. For HEXTE, we also use the standard mode data, 
providing 64 channel count spectra each 16 s covering 10--250 keV range, to 
perform spectral analysis. HEXTE comprises two clusters, cluster A and cluster 
B, each of which consists of four detectors. We only use the data from cluster 
A, because detector 2 of cluster B loses its spectral capability and automatic 
gain control after 1996 
Mar.~6\footnote{http://heasarc.gsfc.nasa.gov/docs/xte/recipes/hexte.html}.
FTOOLS HXTBACK is used to separate the background data from the
source+background data in raw FITS files, then the source and background
HEXTE spectra are extracted, and the deadtimes of these spectra are corrected 
with command HXTDEAD. Finally, the background-subtracted spectra 
of HEXTE are obtained. In order to improve the signal-to-noise ratios
(S/Ns) of HEXTE spectra, we rebin the HEXTE spectra for per new bin to have
S/Ns larger than 1.5. When extracting spectra, Good Time Interval (GTI) files
are applied, which are created by FTOOLS MAKETIME following the criteria: the
offset between the source and telescope pointing direction is less than
$0.02^\circ$ and the elevation above the Earth's limb is greater than
$10^\circ$.

\subsection{The BMC model}

Titarchuk et al. (1996, 1997) and Laurent \& Titarchuk (1999) studied the 
Compton up-scattering of soft X-ray photons in a converging flow onto a compact 
star and suggested that the bulk-motion of the converging flow is significant 
in up-scattering photons and then the bulk-motion Comptonization might be 
responsible for the PL spectra seen in X-ray sources. Using observational data 
from the {\it Compton Gamma Ray Observatory, CGRO} and {\it RXTE}, Shrader \& 
Titarchuk (1998) and Titarchuk \& Seifina (2009) investigated the bulk-motion 
Comptonization in BHXBs and found that this kind of Comptonization does exist 
in BHXBs. In XSPEC, the bulk-motion Comptonization is described by the bmc 
model and expressed as:
\begin{equation}
F(E) = \frac{C_{\rm N}}{1+A}(BB+A\times BB*G).
\end{equation}
This expression consists of two components: one is the injected
blackbody-like spectrum (BB), and the other is $BB*G$, a convolution of BB 
with Green's function $G(E,E_{0})$, which accounts for the observed
Comptonization component. This model includes four parameters: the
temperature of the seed photons $kT_{\rm bb}$, spectral index $\alpha$
(photon index $\Gamma=\alpha+1$), a logarithm of the weighting factor
$log(A)$, and the normalization $C_{\rm N}$. The factor $1/(1+A)$ is the
fraction of the seed photon radiation which is not affected by a noticeable
up-scattering in the plasma cloud and seen directly by the Earth observer,
whereas the Comptonization factor $f$ ($f=A/(1+A)$) is the fraction of the seed
photon radiation which is up-scattered by the Compton cloud. High values of
$\alpha$ indicate the low efficiency of Comptonization that usually occurs
when the thermal equilibrium between the Compton cloud and the seed photon
environment is established. In two cases, the BMC model is reduced to the BB
model: one is with  $A\ll1$, indicating that the convolution Comptonization
component is negligible; the other is with $A\gg1~\&~\alpha\gg1$,
meaning that near all the seed photons are up-scattered by the Compton cloud
and, however the efficiency of Comptonization is small.

\subsection{Spectral fitting}

\begin{figure*}[t]
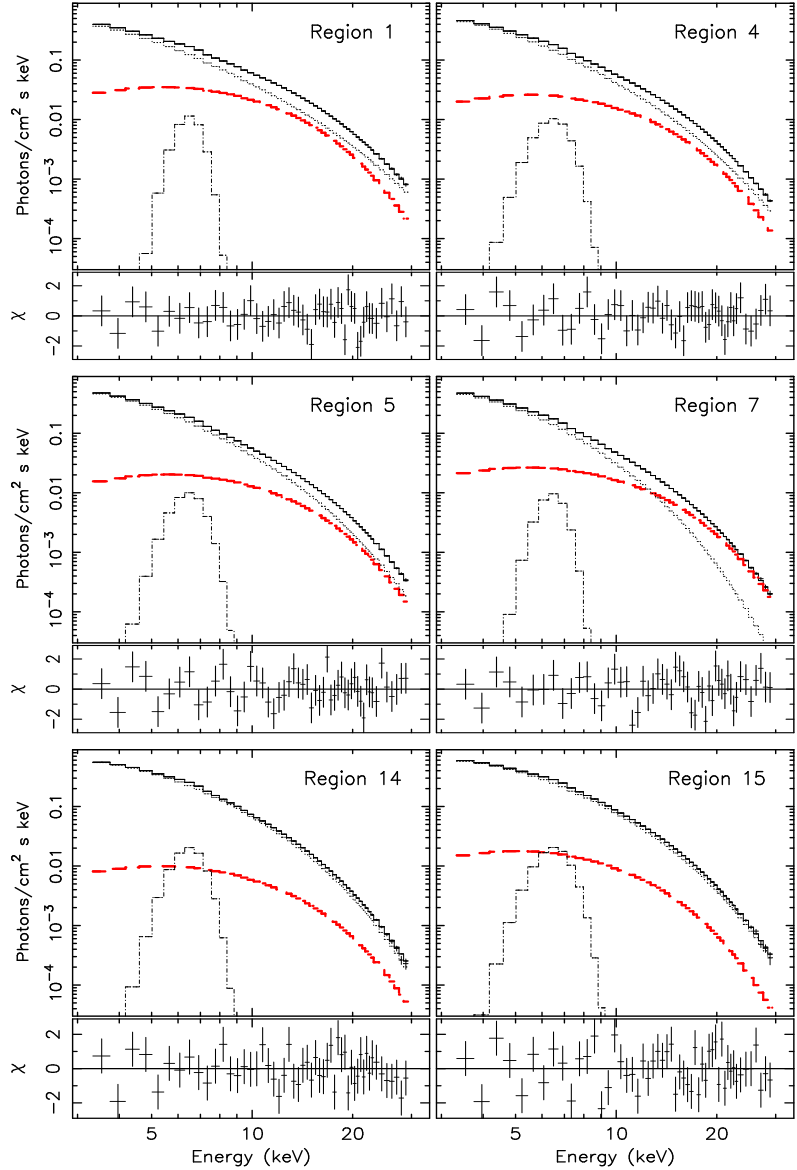

\vspace{1.2cm}
\centerline{\hbox{
\includegraphics[width=3.6cm,height=5.6cm,angle=-90]{f3_1_1.ps}
\hspace{-0.15cm}
\includegraphics[width=3.6cm,height=4.8cm,angle=-90]{f3_2_1.ps}
}}

\vspace{-0.055cm}
\centerline{\hbox{
\includegraphics[width=1.2cm,height=5.6cm,angle=-90]{f3_1_2.ps}
\hspace{-0.15cm}
\includegraphics[width=1.2cm,height=4.8cm,angle=-90]{f3_2_2.ps}
}}

\vspace{0.15cm}
\centerline{\hbox{
\includegraphics[width=3.6cm,height=5.6cm,angle=-90]{f3_3_1.ps}
\hspace{-0.15cm}
\includegraphics[width=3.6cm,height=4.8cm,angle=-90]{f3_4_1.ps}
}}

\vspace{-0.055cm}
\centerline{\hbox{
\includegraphics[width=1.2cm,height=5.6cm,angle=-90]{f3_3_2.ps}
\hspace{-0.15cm}
\includegraphics[width=1.2cm,height=4.8cm,angle=-90]{f3_4_2.ps}
}}

\vspace{0.15cm}
\centerline{\hbox{
\includegraphics[width=3.6cm,height=5.6cm,angle=-90]{f3_5_1.ps}
\hspace{-0.15cm}
\includegraphics[width=3.6cm,height=4.8cm,angle=-90]{f3_6_1.ps}
}}

\vspace{-0.055cm}
\centerline{\hbox{
\includegraphics[width=2.cm,height=5.6cm,angle=-90]{f3_5_2.ps}
\hspace{-0.15cm}
\includegraphics[width=2.cm,height=4.8cm,angle=-90]{f3_6_2.ps}
}}

\vspace{0.25cm}
\caption{Six unfolded spectra of PCA spectral fitting in 3-30 keV, showing 
three individual components, namely, BB (red dashed line), LINE (dot-dashed 
line), and CPL (dotted line). The residuals ($\chi$) in terms of $\sigma$ 
with error bar for each spectral fitting are demonstrated. The residual 
distributions show that these fittings are statistically 
good.}\label{fig:eufs_pca}
\end{figure*}

The Eastern and Western models have been two classical spectral models for NSXBs 
(Mitsuda et al. 1984, 1989; White et al. 1986, 1988). Church \& Ba{\l}uci{\'n}ska-Church 
(1995, 2004) developed the two models into an extended accretion disk corona (ADC) 
model which consists of a blackbody (BB) interpreted as the emission from the NS 
surface and a cut off powerlaw (CPL) representing the Comptonization of the soft 
photons from the disk by the energetic electrons from an extend ADC above the disk. 
The extended ADC model was used to fit the PCA spectra of Z sources (Church et al. 
2006, 2012; Jackson et al. 2009; Ba{\l}uci{\'n}ska-Church et al. 2010). In this work, 
we accept this model. Taking into account the Fe-${\rm K_{\alpha}}$ emission, we use 
a combination model of BB+LINE+CPL to fit the 17 PCA spectra in 3-30 keV. Di Salvo 
et al. (2000) detected an absorption edge at $\sim$8.5 keV in the broadband 
{\it BeppoSAX} spectra of GX 17+2. In our practice, a unusual structure is presented 
in $\sim$(8-10) keV in the residual distribution when we use the BB+LINE+CPL model 
to fit the PCA spectra. If an absorption edge is added into the model, the fitting 
can be improved, so it is also taken into account in our fitting. Certainly, the 
interstellar absorption is taken into consideration and, however, the lack of data 
below $\sim$3~keV prevents XSPEC from determining the photoelectric interstellar 
hydrogen column density $(N_{\rm H})$. But, the values of $(N_{\rm H})$ is larger 
than $3\times10^{22}\ {\rm atom\ cm^{-2}}$ when the extended ADC model is used to fit 
the PCA spectra of Z sources except Sco X-1 (Church et al. 2006, 2012; Jackson et al. 
2009; Ba{\l}uci{\'n}ska-Church et al. 2010), so the PCA spectra is feasible to measure 
the values of $N_{\rm H}$ of these Z sources. Due to the calibration uncertainties, a 
systematic error of 0.5\% is added into the PCA spectra (Barret et al. 2000). The PCA 
response matrixes are created with {\it RXTE} perl script PCARSP and the spectra are 
fitted with XSPEC version 12.7.0. The fitting results are listed in Table~\ref{tab:pca} 
and six unfolded spectra are shown in Figure~\ref{fig:eufs_pca}.                 

\begin{figure*}[t]
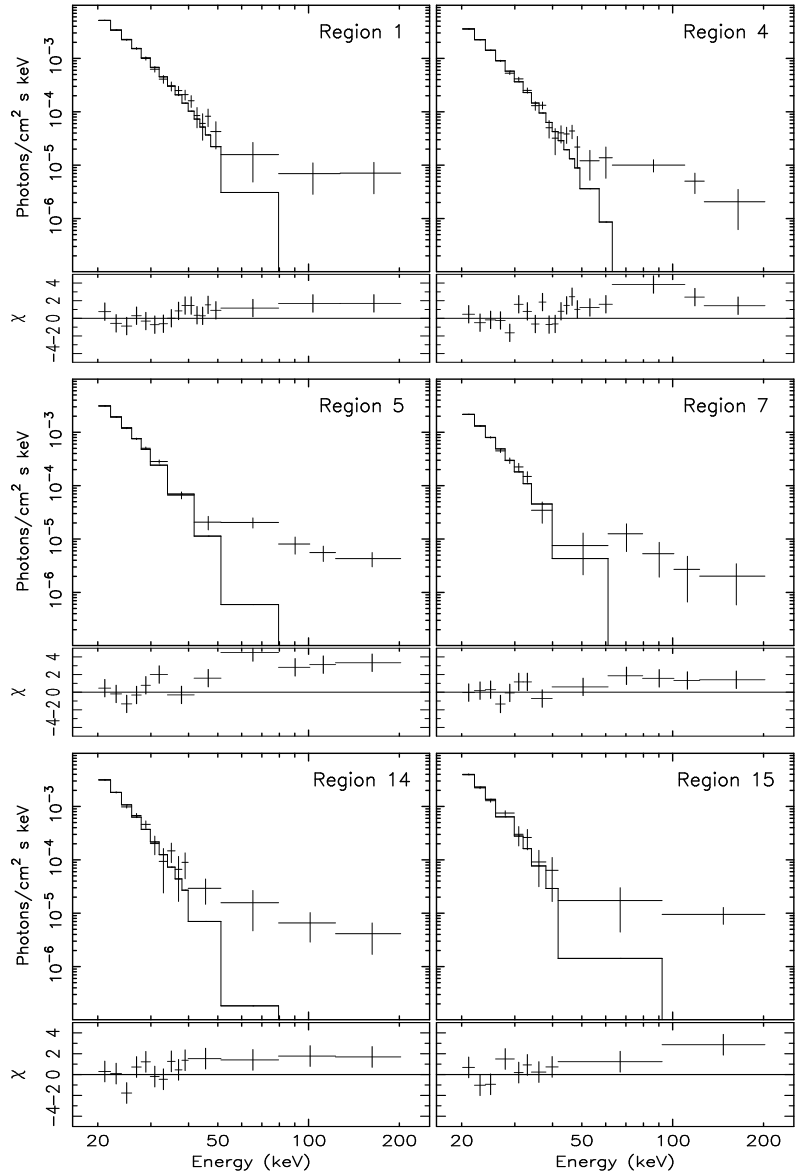

\vspace{1.2cm}
\centerline{\hbox{
\includegraphics[width=3.6cm,height=5.6cm,angle=-90]{f5_1_1.ps}
\hspace{-0.15cm}
\includegraphics[width=3.6cm,height=4.8cm,angle=-90]{f5_2_1.ps}
}}

\vspace{-0.045cm}
\centerline{\hbox{
\includegraphics[width=1.2cm,height=5.6cm,angle=-90]{f5_1_2.ps}
\hspace{-0.15cm}
\includegraphics[width=1.2cm,height=4.8cm,angle=-90]{f5_2_2.ps}
}}

\vspace{0.15cm}
\centerline{\hbox{
\includegraphics[width=3.6cm,height=5.6cm,angle=-90]{f5_3_1.ps}
\hspace{-0.15cm}
\includegraphics[width=3.6cm,height=4.8cm,angle=-90]{f5_4_1.ps}
}}

\vspace{-0.045cm}
\centerline{\hbox{
\includegraphics[width=1.2cm,height=5.6cm,angle=-90]{f5_3_2.ps}
\hspace{-0.15cm}
\includegraphics[width=1.2cm,height=4.8cm,angle=-90]{f5_4_2.ps}
}}

\vspace{0.15cm}
\centerline{\hbox{
\includegraphics[width=3.6cm,height=5.6cm,angle=-90]{f5_5_1.ps}
\hspace{-0.15cm}
\includegraphics[width=3.6cm,height=4.8cm,angle=-90]{f5_6_1.ps}
}}

\vspace{-0.045cm}
\centerline{\hbox{
\includegraphics[width=2.cm,height=5.6cm,angle=-90]{f5_5_2.ps}
\hspace{-0.15cm}
\includegraphics[width=2.cm,height=4.8cm,angle=-90]{f5_6_2.ps}
}}

\vspace{0.25cm}
\caption{Six unfolded spectra of HEXTE spectral fitting in 20-200 
keV, showing an individual component, namely, BREMSS (solid line). 
The residuals ($\chi$) in terms of $\sigma$ with error bar for each 
spectral fitting are demonstrated. Either these unfolded spectra or 
residual distributions show that there are obvious high-energy 
excesses in the energy bands higher than $\sim$40 
keV.}\label{fig:eufs_hexte_2}
\end{figure*}

The HEXTE response matrix of Cluster A (xh97mar20c$_{-}$pwa$_{-}$64b.rmf) and the 
ancillary response file of Cluster A (hexte$_{-}$00may26$_{-}$pwa.arf) provided by 
{\it RXTE} team are applied when the HXTE spectra are fitted. Using a two-component 
model consisting of a thermal bremsstrahlung (BREMSS) and a simple PL, D'Amico et al. 
(2001) fit the HEXTE spectra of Sco X-1 in which a hard tail was detected. In this 
work, we use this two-component model or a single-component model, i.e. the BREMSS 
model, to fit the 17 HEXTE spectra in 20-200 keV. We use the XSPEC convolution model 
CFLUX to calculate the unabsorbed flux of BREMSS in 20-50 keV and the unabsorbed flux 
of PL in 20-200 keV. The fitting results are listed in Table~\ref{tab:hexte} and some 
unfolded spectra are shown in Figures~\ref{fig:eufs_hexte_1}-\ref{fig:eufs_hexte_3}. 

We jointly fit six PCA+HEXTE spectra in 3-200 keV with a combination model 
consisting of a BMC, a LINE, and a CPL. A multiplicative constant is added to the 
spectral model for allowing the HEXTE spectral normalization to float with the PCA 
spectral normalization. In our practice, we fix this constant at 1 for the PCA 
spectra, while it is free for the HEXTE spectra, varying between 1 and 1.3. The 
fitting results are listed in Table~\ref{tab:pca+hexte} and six unfolded spectra 
are shown in Figure~\ref{fig:eufs_pca+hexte}.

\section{Results and discussion}

\subsection{PCA spectral fitting}

The values of ${\rm \chi^2_{\rm \nu}}$ (${\rm \chi^2_{\rm \nu}} = {\rm \chi^2/dof}$, 
refer to Table~\ref{tab:pca}) and the residual distributions shown in 
Figure~\ref{fig:eufs_pca} show that the pca spectral fittings are statistically good. 
Using the same spectral model, i.e. the BB+LINE+CPL model, Church et al. (2012) fit the 
PCA spectra on another ``Z'' track of this source. In their fitting, $\Gamma$, i.e. photon 
index, is fixed at 1.7 throughout the ``Z'' track. In our fitting, we free this 
parameter in order to study its evolution. Our fitting results show that in the HB the 
value of $\Gamma$ is around 1.5, basically being consistent with the fixed value of this 
parameter of Church et al. (2012), while it decreases remarkably in the NB and FB. In 
most cases of the NB and FB, the photon index is so small that its errors are relatively 
large, so we fix it at a relatively small value. Except $\Gamma$, other fitting parameters 
listed in Table~\ref{tab:pca} are consistent with those of Church et al. (2012) in genaral. 
The CPL component in the BB+LINE+CPL model is interpreted as the Comptonized emission of 
the disk soft photons inversely Comptonized by the energetic electrons from an extended 
ADC above the disk, while the BB component is interpreted as the emission from the NS 
surface (Church \& Ba{\l}uci{\'n}ska-Church 1995, 2004). It is noted that the BB temperature 
spans a range of $\sim$(2.4-2.9) keV, as listed in Table~\ref{tab:pca}, which confirm the 
explanation that the BB component in the BB+LINE+CPL model is from the NS, because the 
highest temperature of the disk, i.e. the inner disk temperature, is less than 2 keV 
(Mitsuda et al. 1984, 1989; Cackett et al. 2008; Lin et al. 2007, 2009; 
Ding et al. 2006a, 2011).                

\begin{figure*}[t]
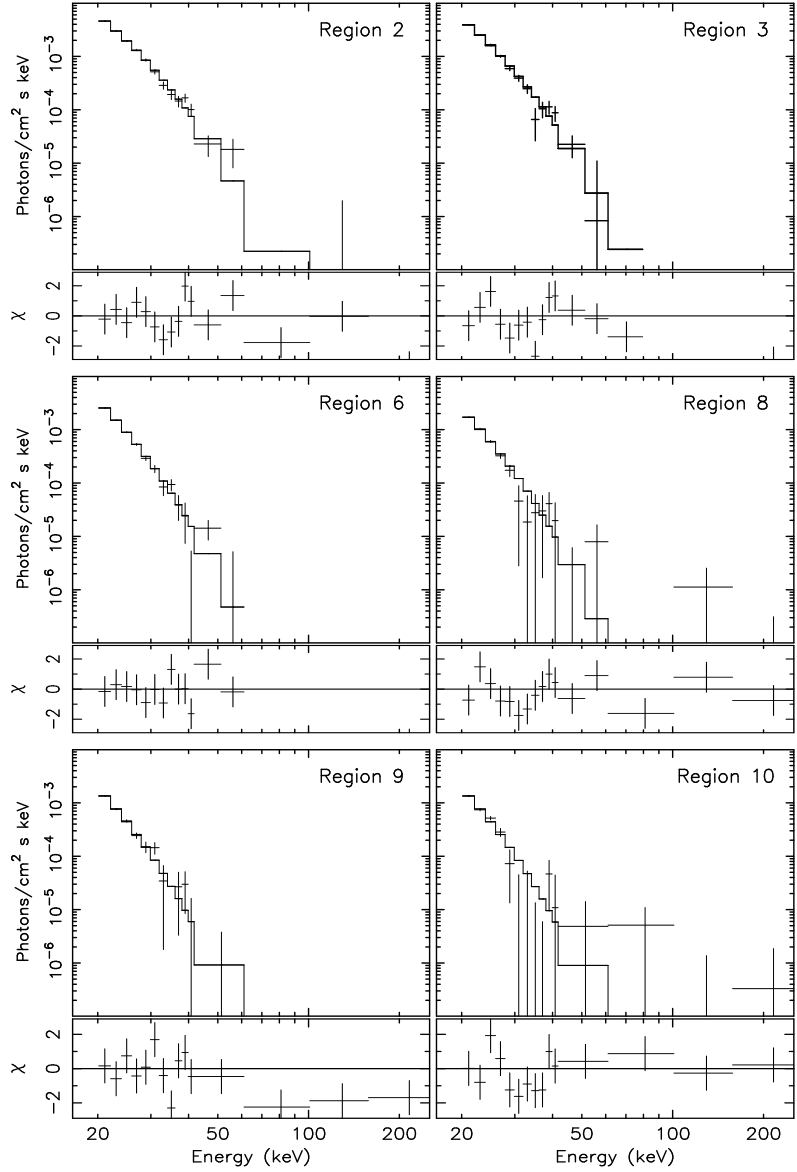

\vspace{1.2cm}
\centerline{\hbox{
\includegraphics[width=3.6cm,height=5.6cm,angle=-90]{f4_1_1.ps}
\hspace{-0.15cm}
\includegraphics[width=3.6cm,height=4.8cm,angle=-90]{f4_2_1.ps}
}}

\vspace{-0.055cm}
\centerline{\hbox{
\includegraphics[width=1.2cm,height=5.6cm,angle=-90]{f4_1_2.ps}
\hspace{-0.15cm}
\includegraphics[width=1.2cm,height=4.8cm,angle=-90]{f4_2_2.ps}
}}

\vspace{0.15cm}
\centerline{\hbox{
\includegraphics[width=3.6cm,height=5.6cm,angle=-90]{f4_3_1.ps}
\hspace{-0.15cm}
\includegraphics[width=3.6cm,height=4.8cm,angle=-90]{f4_4_1.ps}
}}

\vspace{-0.055cm}
\centerline{\hbox{
\includegraphics[width=1.2cm,height=5.6cm,angle=-90]{f4_3_2.ps}
\hspace{-0.15cm}
\includegraphics[width=1.2cm,height=4.8cm,angle=-90]{f4_4_2.ps}
}}

\vspace{0.15cm}
\centerline{\hbox{
\includegraphics[width=3.6cm,height=5.6cm,angle=-90]{f4_5_1.ps}
\hspace{-0.15cm}
\includegraphics[width=3.6cm,height=4.8cm,angle=-90]{f4_6_1.ps}
}}

\vspace{-0.055cm}
\centerline{\hbox{
\includegraphics[width=2.cm,height=5.6cm,angle=-90]{f4_5_2.ps}
\hspace{-0.15cm}
\includegraphics[width=2.cm,height=4.8cm,angle=-90]{f4_6_2.ps}
}}

\vspace{0.25cm}
\caption{Six unfolded spectra of HEXTE spectral fitting in 20-200 
keV, showing an individual component, namely, BREMSS (solid line). 
The residuals ($\chi$) in terms of $\sigma$ with error bar for each 
spectral fitting are demonstrated. The residual distributions show 
that these fittings are statistically 
acceptable.}\label{fig:eufs_hexte_1}
\end{figure*}

\begin{figure*}[t]
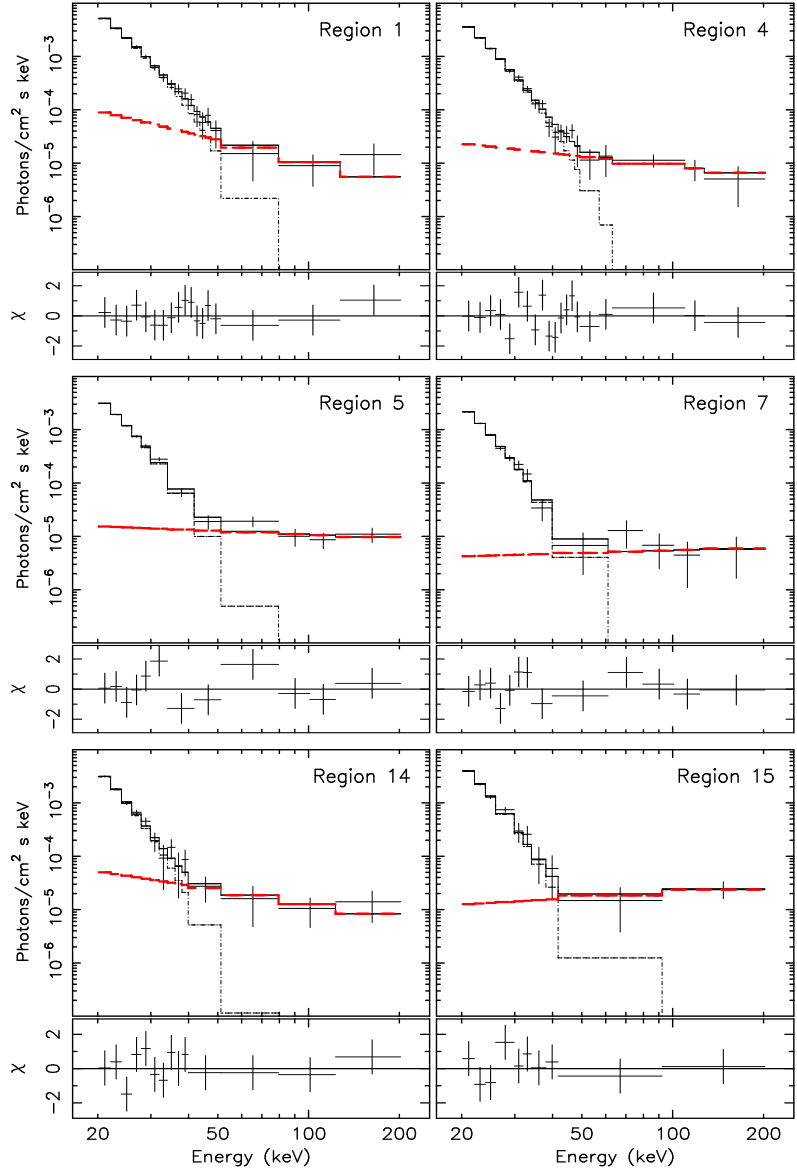

\vspace{1.2cm}
\centerline{\hbox{
\includegraphics[width=3.6cm,height=5.6cm,angle=-90]{f6_1_1.ps}
\hspace{-0.15cm}
\includegraphics[width=3.6cm,height=4.8cm,angle=-90]{f6_2_1.ps}
}}

\vspace{-0.055cm}
\centerline{\hbox{
\includegraphics[width=1.2cm,height=5.6cm,angle=-90]{f6_1_2.ps}
\hspace{-0.15cm}
\includegraphics[width=1.2cm,height=4.8cm,angle=-90]{f6_2_2.ps}
}}

\vspace{0.15cm}
\centerline{\hbox{
\includegraphics[width=3.6cm,height=5.6cm,angle=-90]{f6_3_1.ps}
\hspace{-0.15cm}
\includegraphics[width=3.6cm,height=4.8cm,angle=-90]{f6_4_1.ps}
}}

\vspace{-0.055cm}
\centerline{\hbox{
\includegraphics[width=1.2cm,height=5.6cm,angle=-90]{f6_3_2.ps}
\hspace{-0.15cm}
\includegraphics[width=1.2cm,height=4.8cm,angle=-90]{f6_4_2.ps}
}}

\vspace{0.15cm}
\centerline{\hbox{
\includegraphics[width=3.6cm,height=5.6cm,angle=-90]{f6_5_1.ps}
\hspace{-0.15cm}
\includegraphics[width=3.6cm,height=4.8cm,angle=-90]{f6_6_1.ps}
}}

\vspace{-0.055cm}
\centerline{\hbox{
\includegraphics[width=2.cm,height=5.6cm,angle=-90]{f6_5_2.ps}
\hspace{-0.15cm}
\includegraphics[width=2.cm,height=4.8cm,angle=-90]{f6_6_2.ps}
}}

\vspace{0.25cm}
\caption{Six unfolded spectra of HEXTE spectral fitting in 20-200 keV, 
showing two individual components, namely, BREMSS (dot-dashed line) and 
PL (red dashed line). The residuals ($\chi$) in terms of $\sigma$ with 
error bar for each spectral fitting are demonstrated. The residual 
distributions show that these fittings are statistically 
good.}\label{fig:eufs_hexte_3}
\end{figure*}

\subsection{HEXTE spectral fitting}

we resort to the two-component model of D'Amico et al. (2001) to perform analysis for 
the 17 HEXTE spectra. Firstly, we use a single-component model, i.e. the BREMSS model, 
to fit the 17 HEXTE spectra in 20-200 keV. We find that some HEXTE spectra can be fitted 
statistically well by this single-component model. Six representative unfolded spectra of 
the single-component fitting as well as the corresponding residual distributions are shown 
in Figure~\ref{fig:eufs_hexte_1}. From the residual distributions shown in 
Figure~\ref{fig:eufs_hexte_1} and the corresponding values of ${\rm \chi^2_{\rm \nu}}$ 
(refer to Table~\ref{tab:hexte}), one can conclude that these single-component fitting are 
statistically acceptable. However, when six HEXTE spectra, i.e. the HEXTE spectra of 
HID regions 1, 4, 5, 7, 14, 15, are fitted by the BREMSS model, remarkable high-energy 
excesses in the energy bands above $\sim$40 keV are shown in the unfolded spectra or  
residual distributions, which are demonstrated by Figure~\ref{fig:eufs_hexte_2}. 
Secondly, we use the two-component model, i.e. the BREMSS+PL model, to fit the six HEXTE 
spectra. The fitting parameters are listed in Table~\ref{tab:hexte} and the six unfolded 
spectra of the two-component fitting as well as the corresponding residual distributions 
are shown in Figure~\ref{fig:eufs_hexte_3}. Either the values of $\chi^2_{\rm \nu}$ 
(refer to Table~\ref{tab:hexte}) or the residual distributions shown in 
Figure~\ref{fig:eufs_hexte_3} show that these two-component fittings are statistically 
good. Moreover, comparing the values of ${\rm \chi^2}$ and dof with and without the PL 
component in the spectral model for the six HEXTE spectra, listed in Table~\ref{tab:hexte}, 
one can conclude that the fittings are statistically better with PL than without PL, and, 
what is more, the fittings without the PL component are statistically unacceptable in two 
cases. Furthermore, the F-test probabilities for adding the PL component in the spectra 
model, listed in Table~\ref{tab:hexte}, span a range of $\sim$($10^{-2}$-$10^{-5}$), 
which validate the PL component in these HEXTE spectra. Our analyses suggest that a PL 
hard X-ray tail is detected in each of the six HEXTE spectra of HID regions 1, 4, 5, 
7, 14, 15.   

In any of the three branches throughout the ``Z'' track, the hard tail is detected in 
two regions. As the listed photon indices ($\Gamma$) in Table~\ref{tab:hexte}, the hard 
tail continuously hardens on the HB and NB, but it softens from NB to FB, then it hardens 
again on the FB. The softest and hardest photon indices of the hard tail are obtained on 
the HB and FB, respectively. Except the photon index of region 14, the hard tail becomes 
hard continuously along the ``Z'' track in the sequence HB$\rightarrow$NB$\rightarrow$FB, 
which is consistent with hard tail behavior of Sco X-1 (D'Amico et al. 2001). The detected 
hard tail in region 1 contributes $\sim$20\% of the total flux in the 20-200 keV 
energy interval. Interestingly, the ratio of the PL flux to the total flux continuously 
increases along the ``Z'' track, so that the hard tail dominates half of the total flux 
in region 15.   

\subsection{PCA+HEXTE spectral fitting}

\begin{figure*}[t]
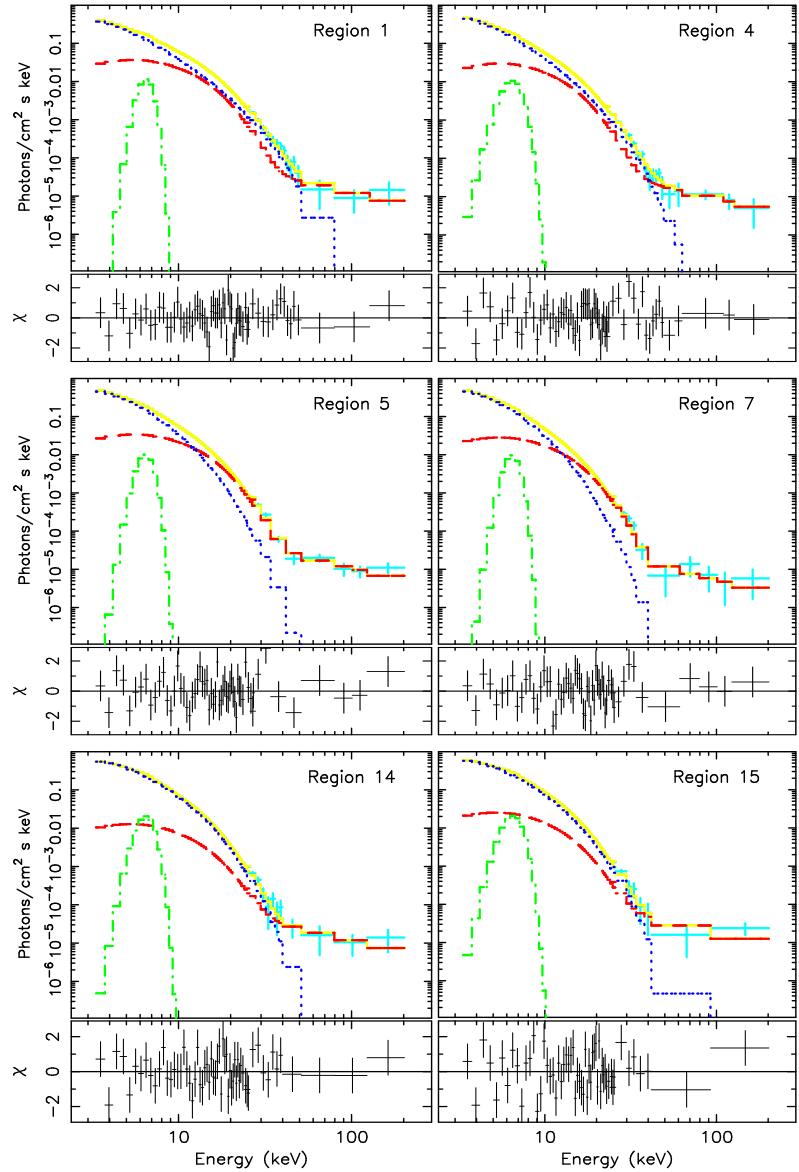

\vspace{1.2cm}
\centerline{\hbox{
\includegraphics[width=3.6cm,height=5.6cm,angle=-90]{f7_1_1.ps}
\hspace{-0.15cm}
\includegraphics[width=3.6cm,height=4.8cm,angle=-90]{f7_2_1.ps}
}}

\vspace{-0.055cm}
\centerline{\hbox{
\includegraphics[width=1.2cm,height=5.6cm,angle=-90]{f7_1_2.ps}
\hspace{-0.15cm}
\includegraphics[width=1.2cm,height=4.8cm,angle=-90]{f7_2_2.ps}
}}

\vspace{0.15cm}
\centerline{\hbox{
\includegraphics[width=3.6cm,height=5.6cm,angle=-90]{f7_3_1.ps}
\hspace{-0.15cm}
\includegraphics[width=3.6cm,height=4.8cm,angle=-90]{f7_4_1.ps}
}}

\vspace{-0.055cm}
\centerline{\hbox{
\includegraphics[width=1.2cm,height=5.6cm,angle=-90]{f7_3_2.ps}
\hspace{-0.15cm}
\includegraphics[width=1.2cm,height=4.8cm,angle=-90]{f7_4_2.ps}
}}

\vspace{0.15cm}
\centerline{\hbox{
\includegraphics[width=3.6cm,height=5.6cm,angle=-90]{f7_5_1.ps}
\hspace{-0.15cm}
\includegraphics[width=3.6cm,height=4.8cm,angle=-90]{f7_6_1.ps}
}}

\vspace{-0.055cm}
\centerline{\hbox{
\includegraphics[width=2.cm,height=5.6cm,angle=-90]{f7_5_2.ps}
\hspace{-0.15cm}
\includegraphics[width=2.cm,height=4.8cm,angle=-90]{f7_6_2.ps}
}}

\vspace{0.25cm}
\caption{Six unfolded spectra of PCA+HEXTE spectral fitting in 3-200 
keV, in which hard tails are detected, showing three individual 
components, namely, BMC (red dashed line), LINE (green dot-dashed line), 
and CPL (blue dotted line), as well as the sum of individual components 
(solid yellow line) and the data (celeste crosses). The residuals 
($\chi$) in terms of $\sigma$ with error bar for each spectral fitting 
are demonstrated. The residual distributions show that these fittings 
are statistically good.}\label{fig:eufs_pca+hexte}
\end{figure*}

In order to investigate the possible mechanism for producing the hard X-ray emission 
in GX 17+2, we jointly fit the six PCA+HEXTE spectra in whose HEXTE spectra the hard 
tails are detected. Replacing the BB component in the BB+LINE+CPL model with the BMC 
component, this tri-component model of PCA spectral fitting turns into another 
tri-component model, i.e. the BMC+LINE+CPL model. We use the BMC+LINE+CPL model to 
fit the six PCA+HEXTE spectra in 3-200 keV. The fitting parameters are listed in 
Table~\ref{tab:pca+hexte} and the six unfolded spectra as well as the corresponding 
residual distributions are shown in Figure~\ref{fig:eufs_pca+hexte}. When fitting, we 
find that the errors of the energy spectral index of BMC ($\alpha$) are large if it is 
free, so we free this parameters at first and then fix it at its steady fitting value. 
Either the values of $\chi^2_{\rm \nu}$ (refer to Table~\ref{tab:pca+hexte}) or the 
residual distributions shown in Figure~\ref{fig:eufs_pca+hexte} show that these 
spectral fittings are statistically good. 

Looking at the parameters listed in Table~\ref{tab:pca+hexte} and Table~\ref{tab:pca}, 
one can see that the LINE or CPL parameters listed in Table~\ref{tab:pca+hexte} are 
consistent with the corresponding LINE or CPL parameters listed in Table~\ref{tab:pca}. 
Significantly, the BB temperatures of the seed photons for Comptonization in the BMC 
process, listed in Table~\ref{tab:pca+hexte}, are consistent with the corresponding BB 
temperatures listed in Table~\ref{tab:pca} very well, indicating that the injected BB 
component in the BMC model is the BB component in the BB+LINE+CPL model exactly. 
Visibly, in each unfolded spectrum in Figure~\ref{fig:eufs_pca+hexte}, the low-energy 
segment of the BMC component (below $\sim$40 keV) displays the injected BB component, 
while high-energy segment of the BMC component (above $\sim$40 keV) shows the observed 
Comptonization component. Looking at the red dashed lines in 
Figure~\ref{fig:eufs_pca+hexte} and the red dashed lines in Figure~\ref{fig:eufs_pca}, 
one can see that the injected BB components in the BMC are just the BB components in 
the BB+LINE+CPL model, while the Comptonization components in the BMC fit the 
high-energy spectra very well. The BB component in the BB+LINE+CPL model is 
interpreted as the emission from the NS surface. Therefore, our analyses suggest that 
the BMC process taking place around the NS could be an alternative mechanism for 
producing the detected hard X-ray tails in GX 17+2.                          

\section{Conclusion}

In this work, using the {\it RXTE} observations for Z source GX 17+2, we study the 
evolution of its PCA spectra and HEXTE spectra along a complete ``Z'' track on its 
HID. In 3-30 keV, the PCA spectra can be fit by the BB+LINE+CPL model statistically 
well. In the HEXTE spectra, a hard X-ray tail is discontinuously detected throughout 
the ``Z'' track. The detected hard tail hardens in the sequence 
HB$\rightarrow$NB$\rightarrow$FB in general. In the 20-200 keV energy interval, the 
hard tail contributes $\sim$(20-50)\% of the total flux. The results of jointly 
fitting the PCA+HEXTE spectra in 3-200 keV with the BMC+LINE+CPL model suggest that 
the BMC process taking place around the NS or in the boundary layer between the NS 
and the disk could be a mechanism for producing the hard X-ray tails.   

\section*{\centering Acknowledgements}

We thank the anonymous referee for her or his constructive comments and suggestions, 
which we have taken to carry out this research deeply. This research has made use of 
the data obtained through the High Energy Astrophysics Science Archive Research Center 
(HEASARC) On-line Service, provided by NASA/Goddard Space Flight Center (GSFC). This 
work is partially supported by National Key Basic Research Program of China (973 
Program 2015CB857100), the Natural Science Foundation of China under grant nos. 
11173024 and 11203064, and the Program of the Light in Chinese Western Region (LCWR) 
under grant no. XBBS 201121 provided by Chinese Academy of Sciences (CAS). This work 
is also partially supported by the 2014 Project of Xinjiang Uygur Autonomous Region 
of China for Flexibly Fetching in Upscale Talents.

\clearpage

\begin{table}
\vspace{3.5cm}
%\footnotesize
\scriptsize
%\tiny
\caption{The spectral fitting parameters of PCA spectra in 3-30 keV for the HID regions of GX 17+2, using a model consisting of a BB, a LINE, and a CPL. Errors quoted are 90\% confidence limits for the fitting parameters ($\Delta\chi^2=2.7$). }\label{tab:pca}
\vspace{0.8mm}
\begin{tabular}{lcccccccccccc}
\hline\hline
 &  &  \multicolumn{2}{c}{BB} & & \multicolumn{3}{c}{LINE} & & \multicolumn{3}{c}{CPL} & \\
          \cline{3-4}  \cline{6-8}  \cline{10-12}      \\
HID    & $^{a}N_{H}$ & $kT_{\rm bb}$ & $N_{\rm bb}$ & & $E_{\rm Fe}$ & $^{b}EW$ & $N_{\rm Fe}$ & & $^{c}\Gamma$& $E_{\rm cut}$ & $N_{\rm cpl}$ & $\chi^2~(dof)$ \\
(No.)  & $(\times10^{-22})$ & (keV) & & & (keV) & (eV) & $(\times10^{-2})$ & & & (keV) & &  \\
\hline
HB \\
1 &  $3.22_{-0.42}^{+0.46}$  &  $2.80_{-0.06}^{+0.09}$ & $7.5_{-1.5}^{+1.6}$ & 
  &  $6.48_{-0.05}^{+0.05}$  &            96           & $1.6_{-0.2}^{+0.3}$ & 
  &  $1.57_{-0.17}^{+0.19}$  &  $7.4_{-0.8}^{+1.0}$    & $5.9_{-1.0}^{+1.4}$ & 34.3(42) \\
2 &  $3.15_{-0.32}^{+0.44}$  &  $2.82_{-0.08}^{+0.09}$ & $6.1_{-1.1}^{+1.6}$ & 
  &  $6.48_{-0.05}^{+0.05}$  &            83           & $1.4_{-0.2}^{+0.3}$ & 
  &  $1.42_{-0.13}^{+0.18}$  &  $6.3_{-0.5}^{+0.8}$    & $5.6_{-0.7}^{+0.8}$ & 43.8(42) \\
3 &  $4.18_{-0.42}^{+0.53}$  &  $2.72_{-0.08}^{+0.09}$ & $7.5_{-1.4}^{+1.9}$ & 
  &  $6.44_{-0.08}^{+0.07}$  &            115          & $2.1_{-0.4}^{+0.7}$ & 
  &  $1.63_{-0.17}^{+0.22}$  &  $6.3_{-0.6}^{+0.9}$    & $8.7_{-1.5}^{+2.4}$ & 45.3(42) \\
4 &  $4.08_{-0.35}^{+0.33}$  &  $2.74_{-0.07}^{+0.10}$ & $6.0_{-1.0}^{+1.2}$ & 
  &  $6.46_{-0.07}^{+0.06}$  &            101          & $1.9_{-0.4}^{+0.4}$ & 
  &  $1.46_{-0.14}^{+0.13}$  &  $5.4_{-0.5}^{+0.4}$    & $7.9_{-1.1}^{+1.3}$ & 37.5(42) \\
\hline
 NB \\
5 &  $3.78_{-0.50}^{+0.37}$  &  $2.86_{-0.13}^{+0.11}$ & $4.2_{-0.5}^{+0.9}$ & 
  &  $6.46_{-0.07}^{+0.06}$  &            92           & $1.8_{-0.4}^{+0.5}$ & 
  &  $1.19_{-0.30}^{+0.16}$  &  $4.5_{-0.8}^{+0.5}$    & $6.6_{-1.4}^{+1.2}$ & 44.7(42) \\
6 &  $3.66_{-0.62}^{+0.49}$  &  $2.88_{-0.07}^{+0.02}$ & $4.8_{-1.1}^{+2.7}$ & 
  &  $6.46_{-0.06}^{+0.06}$  &            88           & $1.6_{-0.3}^{+0.4}$ & 
  &  $0.95_{-0.40}^{+0.28}$  &  $3.5_{-0.8}^{+0.8}$    & $6.0_{-1.5}^{+1.5}$ & 37.5(42) \\
7 &  $3.08_{-0.13}^{+0.13}$  &  $2.83_{-0.03}^{+0.03}$ & $5.5_{-0.5}^{+0.5}$ & 
  &  $6.44_{-0.05}^{+0.05}$  &            77           & $1.4_{-0.2}^{+0.3}$ & 
  &         0.60(fixed)      &  $2.8_{-0.1}^{+0.1}$    & $4.5_{-0.2}^{+0.2}$ & 46.5(43) \\
8 &  $3.07_{-0.13}^{+0.13}$  &  $2.79_{-0.04}^{+0.03}$ & $5.2_{-0.5}^{+0.6}$ & 
  &  $6.48_{-0.05}^{+0.04}$  &            109          & $1.7_{-0.2}^{+0.3}$ & 
  &         0.50(fixed)      &  $2.6_{-0.1}^{+0.1}$    & $4.3_{-0.2}^{+0.2}$ & 44.7(43) \\
9 &  $3.47_{-0.76}^{+0.78}$  &  $2.72_{-0.09}^{+0.13}$ & $4.7_{-2.4}^{+2.2}$ & 
  &  $6.50_{-0.04}^{+0.03}$  &            172          & $2.5_{-0.3}^{+0.4}$ & 
  &  $0.61_{-0.45}^{+0.48}$  &  $2.6_{-0.5}^{+0.8}$    & $4.8_{-1.5}^{+2.1}$ & 41.1(42) \\
\hline
 FB \\
10 &  $3.16_{-0.15}^{+0.14}$ &  $2.78_{-0.06}^{+0.06}$ & $3.7_{-0.7}^{+0.7}$  & 
   &  $6.51_{-0.04}^{+0.03}$ &            193          & $2.9_{-0.3}^{+0.3}$  & 
   &        0.50(fixed)      &  $2.6_{-0.1}^{+0.1}$    & $4.1_{-0.2}^{+0.2}$  & 66.4(43) \\
11 &  $3.24_{-0.14}^{+0.14}$ &  $2.77_{-0.05}^{+0.05}$ & $4.3_{-0.7}^{+0.9}$  & 
   &  $6.49_{-0.04}^{+0.03}$ &            189          & $3.2_{-0.3}^{+0.3}$  & 
   &        0.50(fixed)      &  $2.7_{-0.1}^{+0.1}$    & $4.1_{-0.2}^{+0.2}$  & 52.0(43) \\
12 &  $3.25_{-0.14}^{+0.14}$ &  $2.85_{-0.06}^{+0.07}$ & $3.5_{-0.8}^{+0.8}$  & 
   &  $6.52_{-0.04}^{+0.03}$ &            169          & $3.2_{-0.3}^{+0.3}$  & 
   &        0.50(fixed)      &  $2.9_{-0.1}^{+0.1}$    & $4.0_{-0.2}^{+0.2}$  & 63.6(43) \\
13 &  $3.27_{-0.14}^{+0.15}$ &  $2.80_{-0.06}^{+0.06}$ & $4.1_{-1.0}^{+1.4}$  & 
   &  $6.51_{-0.04}^{+0.04}$ &            166          & $3.5_{-0.3}^{+0.4}$  & 
   &        0.50(fixed)      &  $3.1_{-0.1}^{+0.1}$    & $4.0_{-0.2}^{+0.2}$  & 59.6(43) \\
14 &  $3.13_{-0.15}^{+0.15}$ &  $2.75_{-0.16}^{+0.07}$ & $2.2_{-1.4}^{+1.6}$  & 
   &  $6.48_{-0.05}^{+0.04}$ &            142          & $3.5_{-0.4}^{+0.5}$  & 
   &        0.50(fixed)      &  $3.5_{-0.1}^{+0.1}$    & $3.8_{-0.2}^{+0.2}$  & 40.3(43) \\
15 &  $3.22_{-0.17}^{+0.18}$ &  $2.50_{-0.36}^{+0.16}$ & $4.9_{-1.1}^{+1.5}$  & 
   &  $6.51_{-0.06}^{+0.06}$ &            141          & $3.9_{-0.7}^{+0.8}$  & 
   &        0.50(fixed)      &  $3.7_{-0.2}^{+0.2}$    & $3.8_{-0.2}^{+0.2}$  & 63.5(43) \\
16 &  $2.92_{-0.17}^{+0.16}$ &  $2.42_{-0.10}^{+0.09}$ & $12.7_{-1.3}^{+1.8}$ & 
   &  $6.47_{-0.06}^{+0.06}$ &            120          & $3.7_{-0.6}^{+0.7}$  & 
   &        0.50(fixed)      &  $3.9_{-0.1}^{+0.1}$    & $3.6_{-0.2}^{+0.2}$  & 44.6(43) \\
17 &  $2.92_{-0.19}^{+0.19}$ &  $2.49_{-0.06}^{+0.06}$ & $21.9_{-1.4}^{+1.5}$ & 
   &  $6.51_{-0.07}^{+0.07}$ &            84           & $2.8_{-0.4}^{+0.5}$  & 
   &        0.50(fixed)      &  $4.0_{-0.2}^{+0.1}$    & $3.7_{-0.2}^{+0.2}$  & 29.9(43) \\
\hline
\end{tabular}
\vspace{-0.5mm}
\\
\footnotesize{
$^{a}${The interstellar hydrogen column density.} \\
$^{b}${The equivalent width of the line component.} \\
$^{c}${The photon index of the CPL component.} \\
}
\end{table}

\clearpage

\begin{table}
\vspace{3.5cm}
\footnotesize
%\scriptsize
%\tiny
\caption{The spectral fitting parameters of the HEXTE spectra in 20-200 keV, using a two-component model consisting of a BREMSS, plus a PL, or a one-component model of BREMSS. Errors quoted are 90\% confidence limits for the fitting parameters ($\Delta\chi^2=2.7$).}\label{tab:hexte}
\vspace{0.8mm}
\begin{tabular}{lcccccccccc}
\hline\hline
 & & \multicolumn{2}{c}{BREMSS} & & \multicolumn{2}{c}{PL} & & \\
     \cline{3-4} \cline{6-7} \\
 HID (No.) & & $kT$ (keV) & $^{a}$Flux & & $\Gamma$ & $^{b}$Flux & & $\chi^2~(dof)$ & $^{c}$$\chi^2~(dof)$ & $^{d}F$-test  \\
\hline
 HB \\
1   & & $6.28_{-0.74}^{+0.44}$ & $1.17_{-0.43}^{+0.28}$ & & $1.37_{-0.62}^{+0.58}$  & $0.32_{-0.12}^{+0.21}$ & & 6.19(14)  & 18.08(16) & $5.5\times10^{-4}$  \\
2   & & $6.24_{-0.19}^{+0.19}$ & $1.05_{-0.02}^{+0.02}$ & &  -----  &  -----  & & 12.71(11)  &  -----  &  -----  \\
3   & & $5.84_{-0.21}^{+0.22}$ & $0.85_{-0.02}^{+0.02}$ & &  -----  &  -----  & & 19.04(12)  &  -----  &  -----  \\
4   & & $5.50_{-0.17}^{+0.16}$ & $0.74_{-0.02}^{+0.02}$ & & $0.60_{-0.45}^{+0.52}$  & $0.26_{-0.08}^{+0.09}$ & & 14.87(16) & 47.60(18) & $9.1\times10^{-5}$  \\
\hline
 NB \\
5   & & $5.33_{-0.17}^{+0.17}$ & $0.63_{-0.02}^{+0.02}$ & & $0.22_{-0.18}^{+0.15}$  & $0.33_{-0.09}^{+0.09}$ & & 10.58(8)  & 58.51(10) & $1.1\times10^{-3}$  \\
6   & & $4.75_{-0.15}^{+0.15}$ & $0.48_{-0.02}^{+0.02}$ & &  -----  &  -----  & & 8.97(11)   &  -----  &  -----  \\
7   & & $5.02_{-0.26}^{+0.26}$ & $0.42_{-0.02}^{+0.02}$ & & $-0.16_{-0.12}^{+0.14}$ & $0.18_{-0.11}^{+0.11}$ & & 7.00(9)   & 15.16(11) & $3.1\times10^{-2}$  \\
8   & & $4.67_{-0.30}^{+0.32}$ & $0.32_{-0.02}^{+0.02}$ & &  -----  &  -----  & & 11.55(11)  &  -----  &  -----  \\
9   & & $4.41_{-0.29}^{+0.32}$ & $0.24_{-0.02}^{+0.02}$ & &  -----  &  -----  & & 10.95(11)  &  -----  &  -----  \\
\hline
 FB \\
10  & & $4.40_{-0.48}^{+0.56}$ & $0.24_{-0.02}^{+0.02}$ & &  -----  &  -----  & & 14.95(13)  &  -----  &  -----  \\
11  & & $4.46_{-0.39}^{+0.44}$ & $0.30_{-0.02}^{+0.02}$ & &  -----  &  -----  & & 12.39(13)  &  -----  &  -----  \\
12  & & $4.14_{-0.30}^{+0.33}$ & $0.33_{-0.02}^{+0.02}$ & &  -----  &  -----  & & 16.72(12)  &  -----  &  -----  \\
13  & & $4.12_{-0.24}^{+0.26}$ & $0.40_{-0.02}^{+0.02}$ & &  -----  &  -----  & & 8.64(10)   &  -----  &  -----  \\
14  & & $4.35_{-0.51}^{+0.42}$ & $0.54_{-0.09}^{+0.04}$ & & $0.89_{-0.73}^{+0.65}$  & $0.36_{-0.19}^{+0.24}$ & & 7.27(10)  & 19.51(12) & $7.2\times10^{-3}$  \\
15  & & $4.57_{-0.50}^{+0.49}$ & $0.72_{-0.07}^{+0.05}$ & & $-0.33_{-0.30}^{+0.28}$ & $0.72_{-0.39}^{+0.38}$ & & 5.31(6)   & 15.83(8)  & $3.8\times10^{-2}$  \\
16  & & $4.45_{-0.28}^{+0.31}$ & $0.93_{-0.04}^{+0.04}$ & &  -----  &  -----  & & 14.78(12)  &  -----  &  -----  \\
17  & & $4.91_{-0.39}^{+0.43}$ & $1.29_{-0.07}^{+0.07}$ & &  -----  &  -----  & & 5.93(12)   &  -----  &  -----  \\
\hline
\end{tabular}
\vspace{1.mm}
\\
\footnotesize{
$^{a}${The unabsorbed flux in the 20-50 keV range in units of $10^{-9}$ ergs cm$^{-1}$ s$^{-1}$} \\
$^{b}${The unabsorbed flux in the 20-200 keV range in units of $10^{-9}$ ergs cm$^{-1}$ s$^{-1}$} \\
$^{c}${The $\chi^2$ and degree of freedom when the PL component is not included in the spectral model} \\
$^{d}${The probability of chance improvement when a PL is included in the spectral model} \\
}
\end{table}

\clearpage

\begin{table}
\vspace{3.5cm}
%\footnotesize
%\scriptsize
\tiny
\caption{The spectral fitting parameters of the PCA+HEXTE spectra in which a hard tail is detected. The spectra are fit in 3-200 keV, with the model consisting of a BMC, a LINE, and a CPL. Errors quoted are 90\% confidence limits for the fitting parameters ($\Delta\chi^2=2.7$). }\label{tab:pca+hexte}
\vspace{0.8mm}
%\begin{tabular}{lccccccccccccccc}
%\begin{tabular}{p{1.mm}cccccccccccccp{0.01mm}c}
\begin{tabular}{p{1.5mm}p{6.2mm}ccccccccccccp{0.3mm}c}
\hline\hline
 &  &  \multicolumn{4}{c}{BMC} & & \multicolumn{3}{c}{LINE} & & \multicolumn{3}{c}{CPL} & \\
          \cline{3-6}  \cline{8-10}  \cline{12-14}      \\
HID    & $^{a}N_{H}$ & $kT_{\rm bb}$ &  $^{b}\alpha$ & $log{\rm (A)}$ & $N_{\rm bmc}$ & & $E_{\rm Fe}$ & $^{c}EW$ & $N_{\rm Fe}$ & & $^{d}\Gamma$& $E_{\rm cut}$ & $N_{\rm cpl}$ & $^{e}f$ & $\chi^2~(dof)$ \\
(No.)  & & (keV) & & & $(\times10^{-2})$ & & (keV) & (eV) & $(\times10^{-2})$ & & & (keV) & &  \\
\hline
\scriptsize{HB} \\
1 &  $3.3_{-0.4}^{+0.5}$      &  $2.8_{-0.1}^{+0.1}$ & 0.01(fixed)
  &  $-0.58_{-0.31}^{+0.18}$  &  $7.8_{-0.9}^{+0.9}$ & 
  &  $6.5_{-0.1}^{+0.1}$      &           97         & $1.6_{-0.2}^{+0.3}$ & 
  &  $1.6_{-0.2}^{+0.2}$      &  $7.4_{-0.9}^{+1.2}$ & $6.1_{-1.1}^{+1.7}$ & 21\% & 39.4(52) \\
4 &  $4.4_{-0.4}^{+0.5}$      &  $2.7_{-0.1}^{+0.1}$ & 0.01(fixed)
  &  $-0.64_{-0.16}^{+0.12}$  &  $5.8_{-0.5}^{+0.6}$ & 
  &  $6.5_{-0.1}^{+0.1}$      &           111        & $2.1_{-0.4}^{+0.6}$ & 
  &  $1.6_{-0.2}^{+0.2}$      &  $5.8_{-0.5}^{+0.7}$ & $9.1_{-1.4}^{+2.2}$ & 19\% & 57.7(53) \\
\hline
\scriptsize{NB} \\
5 &  $3.2_{-0.2}^{+0.2}$      &  $2.9_{-0.1}^{+0.1}$ & 0.01(fixed)
  &  $-0.69_{-0.10}^{+0.08}$  &  $7.2_{-0.5}^{+0.5}$ & 
  &  $6.5_{-0.1}^{+0.1}$      &           81         & $1.5_{-0.2}^{+0.3}$ & 
  &         0.8(fixed)        &  $3.2_{-0.1}^{+0.1}$ & $4.9_{-0.2}^{+0.2}$ & 17\% & 56.6(48) \\
7 &  $3.1_{-0.2}^{+0.2}$      &  $2.8_{-0.1}^{+0.1}$ & 0.01(fixed)
  &  $-0.88_{-0.29}^{+0.17}$  &  $5.4_{-0.5}^{+0.5}$ & 
  &  $6.4_{-0.1}^{+0.1}$      &           80         & $1.4_{-0.2}^{+0.3}$ & 
  &         0.6(fixed)        &  $2.8_{-0.1}^{+0.1}$ & $4.6_{-0.2}^{+0.2}$ & 12\% & 53.4(48) \\
\hline
\scriptsize{FB} \\
14 &  $3.1_{-0.2}^{+0.2}$      &  $2.7_{-0.2}^{+0.1}$ & 0.01(fixed)
   &  $-0.15_{-0.28}^{+0.33}$  &  $3.4_{-0.8}^{+0.9}$ & 
   &  $6.5_{-0.1}^{+0.1}$      &           142        & $3.5_{-0.4}^{+0.5}$ & 
   &         0.5(fixed)        &  $3.5_{-0.2}^{+0.2}$ & $3.8_{-0.1}^{+0.2}$ & 41\% & 46.2(49) \\
15 &  $3.3_{-0.2}^{+0.2}$      &  $2.6_{-0.2}^{+0.1}$ & 0.01(fixed)
   &  $-0.27_{-0.28}^{+0.20}$  &  $5.5_{-1.2}^{+1.5}$ & 
   &  $6.5_{-0.1}^{+0.1}$      &           145        & $4.0_{-0.7}^{+0.9}$ & 
   &         0.5(fixed)        &  $3.6_{-0.2}^{+0.2}$ & $3.9_{-0.2}^{+0.3}$ & 35\% & 67.4(45) \\
\hline
\end{tabular}
\vspace{-0.5mm}
\\
\footnotesize{
$^{a}${The interstellar hydrogen column density, in units of $10^{-22}\ {\rm cm^2}$.} \\
$^{b}${The energy spectral index of BMC ($\Gamma_{1}=\alpha+1$)} \\
$^{c}${The equivalent width of the line component.} \\
$^{d}${The photon index of the CPL component.} \\
$^{e}${The Comptonization factor of BMC: $f=$A/(A+1)} \\
}
\end{table}

\clearpage

\clearpage

\clearpage

\clearpage

\clearpage

\clearpage

\clearpage

\begin{thebibliography}{}

\bibitem{}Ba{\l}uci{\'n}ska-Church, M., Gibiec, A., Jackson, N. K., and Church, M. J., 2010, \newblock {\it Astron. Astrophys.}, {\bf 512}, A9. 

\bibitem{}Barret, D., Olive, J.~F., Boirin, L., Done, C., Skinner, G.~K., and Grindlay, J.~E., 2000, \newblock {\it Astrophys. J.}, {\bf 533}, 329.

\bibitem{}Cackett, E. M., Miller, J. M., Bhattacharyya, S., Grindlay, J. E., Homan, J., van der Klis, M., Miller, M. C., Strohmayer, T. E., and Wijnands, R., 2008, \newblock {\it Astrophys. J.}, {\bf 674}, 415.

\bibitem{}Church, M. J. and Ba{\l}uci{\'n}ska-Church, M., 1995, \newblock {\it Astron. Astrophys.}, {\bf 300}, 441. 

\bibitem{}Church, M. J. and Ba{\l}uci{\'n}ska-Church, M., 2004, \newblock {\it Mon. Not. R. Astron. Soc.}, {\bf 348}, 955.

\bibitem{}Church, M. J., Gibiec, A., Ba{\l}uci{\'n}ska-Church, M., and Jackson, N. K., 2012, \newblock {\it Astron. Astrophys.}, {\bf 546}, A35.

\bibitem{}Church, M. J., Halai, G. S., and Ba{\l}uci{\'n}ska-Church, M., 2006, \newblock {\it Astron. Astrophys.}, {\bf 460}, 233.

\bibitem{}Coppi, P. S., 1999, in Poutanen J., Svensson R., editors, ASP Conf. Ser. Vol. 161, {\it High Energy Processes in Accreting Black Holes}, Page 375, Astron. Soc. Pac., San Francisco. 

\bibitem{}D'Amico, F., Heindl, W.~A., Rothschild, R.~E., and Gruber, D.~E., 2001, \newblock {\it Astrophys. J.}, {\bf 547}, L147

\bibitem{}D'A{\'i}, A., {\.Z}ycki, P., Di Salvo, T., Iaria, R., Lavagetto, G., and Robba, N. R., 2007, \newblock {\it Astrophys. J.}, {\bf 667}, 411

\bibitem{}Ding, G.~Q., Qu, J.~L., and Li, T.~P., 2003, \newblock {\it Astrophys. J.}, {\bf 596}, L219.

\bibitem{}Ding, G.~Q., Qu, J.~L., and Li, T.~P., 2006a, \newblock {\it Astron. J}, {\bf 131}, 1693.

\bibitem{}Ding, G.~Q., Zhang, S.~N., Li, T.~p., and Qu, J.~L., 2006b, \newblock {\it Astrophysics. J.}, {\bf 654}, 576.

\bibitem{}Ding, G.~Q., Zhang, S.~N., Wang, N., Qu, J.~L., and Yan, S.~P., 2011, \newblock {\it Astron. J}, {\bf 142}, 34.

\bibitem{}Di Salvo, T., Farinelli, R., Burderi, L., Frontera, F., Kuulkers, E., Masetti, N., Robba, N.~R., Stella, L., and van der Klis, M., 2002, \newblock {\it Astron. Astrophys.}, {\bf 386}, 535.

\bibitem{}Di Salvo, T., Goldoni, P., Stella, L., van der Klis, M., Bazzano, A., Burderi, L., Farinelli, R., Frontera, F., Israel, G. L., M{\'e}ndez, M., and 5 coauthors, 2006, \newblock {\it Astrophys. J.}, {\bf 649}, L91.

\bibitem{}Di Salvo, T., Robba, N. R., Iaria, R., Stella, L., Burderi, L., and Israel, G. L., 2001, \newblock {\it Astrophys. J.}, {\bf 554}, 49.

\bibitem{}Di Salvo, T., Stella, L., Robba, N.~R., van der Klis, M., Burderi, L., Israel, G.~L., Homan, J., Campana, S., Frontera, F., and Parmar, A.~N., 2000, \newblock {\it Astrophys. J.}, {\bf 544}, L119.

\bibitem{}Done, C., Gierli{\'n}ski, M., and Kubota, A., 2007, \newblock {\it Astron. Astrophys. Rev.}, {\bf 15}, 1 

\bibitem{}Farinelli, R., Paizis, A., Landi, R., and Titarchuk, L., 2009, \newblock {\it Astron. Astrophys.}, {\bf 498}, 509.

\bibitem{}Farinelli, R., Titarchuk, L., and Frontera, F., 2007, \newblock {\it Astrophys. J.}, {\bf 662}, 1167.                     

\bibitem{}Farinelli, R., Titarchuk, L., Paizis, A., and Frontera, F., 2008, \newblock {\it Astrophys. J.}, {\bf 680}, 602.

\bibitem{}Fender, R.~P., Dahlem, M., Homan, J., Corbel, S., Sault, R., and Belloni, T.~M., 2007, \newblock {\it Mon. Not. R. Astron. Soc.}, {\bf 380}, L25.

\bibitem{}Fiocchi M., Bazzano A., Ubertini P., and Jean P., 2006, \newblock {\it Astrophys. J.}, {\bf 651}, 416.

\bibitem{}Focke, W. B., 1996, \newblock {\it Astrophys. J.}, {\bf 470}, L127.

\bibitem{}Gierli{\'n}ski, M., Zdziarski, A.~A., Poutanen, J., Coppi, P.~S., Ebisawa, K., and Johnson, W.~N., 1999, \newblock {\it Mon. Not. R. Astron. Soc.}, {\bf 309}, 496.

\bibitem{}Hasinger, G., van der Klis, 1989, \newblock {\it Astron. Astrophys.}, {\bf 225}, 79.

\bibitem{}Homan, J., van der Klis, M., Jonker, P.~G., Wijnands, R., Kuulkers, E., M{\'e}ndez, M., and Lewin, W.~H.~G., 2002, \newblock {\it Astrophys. J.}, {\bf 568}, 878.

\bibitem{}Iaria, R., Burderi, L., Di Salvo, T., La Barbera, A., and Robba, N. R., 2001, \newblock {\it Astrophys. J.}, {\bf 547}, 412.

\bibitem{}Iaria, R., Di Salvo, T., Robba, N.~R., and Burderi, L., 2002, \newblock {\it Astrophys. J.}, {\bf 567}, 503.

\bibitem{}Jackson, N. K., Church, M. J., and Ba{\l}uci{\'n}ska-Church, M., 2009, \newblock {\it Astron. Astrophys.}, {\bf 494}, 1059. 

\bibitem{}Lavagetto, G., Iaria, R., Di Salvo, T., Burderi, L., Robba, N. R., Frontera, F., and Stella, L., 2004, \newblock {\it NuPhS}, {\bf 132}, 616.

\bibitem{}Laurent, P. and Titarchuk, L., 1999, \newblock {\it Astrophys. J.}, {\bf 511}, 289.

\bibitem{}Lin, D., Remillard, R.~A., and Homan, J., 2007, \newblock {\it Astrophys. J.}, {\bf 667}, 1073.

\bibitem{}Lin, D., Remillard, R.~A., and Homan, J., 2009, \newblock {\it Astrophys. J.}, {\bf 696}, 1257. 

\bibitem{}Migliari, S., Miller-Jones, J.~C.~A., Fender, R.~P., Homan, J., Di Salvo, T., Rothschild, R.~E., Rupen, M.~P., Tomsick, J.~A., Wijnands, R., and van der Klis, M., 2007, \newblock {\it Astrophys. J.}, {\bf 671}, 706. 

\bibitem{}Mitsuda, K., Inoue, H., Koyama, K., Makishima, K., Matsuoka, M., Ogawara, Y., Suzuki, K., Tanaka, Y., Shibazaki, N., and Hirano, T., 1984, \newblock {\it PASJ}, {\bf 36}, 741. 

\bibitem{}Mitsuda, K., Inoue, H., Nakamura, N., and Tanaka, Y., 1989, \newblock {\it PASJ}, {\bf 41}, 97.

\bibitem{}Paizis, A., Farinelli, R., Titarchuk, L., Courvoisier, T.~J.-L., Bazzano, A., Beckmann, V., Frontera, F., Goldoni, P., Kuulkers, E., Mereghetti, S., Rodriguez, J., and Vilhu, O., 2006, \newblock {\it Astron. Astrophys.}, {\bf 459}, 187.

\bibitem{}Penninx, W., Lewin, W.~H.~G., Zijlstra, A.~A., Mitsuda, K., and van Paradijs, J., 1988, \newblock {\it Nature}, {\bf 336}, 146. 

\bibitem{}Piraino, S., Santangelo, A., di Salvo, T., Kaaret, P., Horns, D., Iaria, R., and Burderi, L., 2007, \newblock {\it Astron. Astrophys.}, {\bf 471}, L17.

\bibitem{}Piraino S., Santangelo A., Ford E. C., and Kaaret P., 1999, \newblock {\it Astron. Astrophys.}, {\bf 349}, L77.

\bibitem{}Poutanen, J. and Coppi, P.~S., 1998, \newblock {\it Phys. Scr.}, {\bf T77}, 57. 

\bibitem{}Raichur, H., Misra, R., and Dewangan, G., 2011, \newblock {\it Mon. Not. R. Astron. Soc.}, {\bf 416}, 637.

\bibitem{}Revnivtsev, M. G., Tsygankov, S. S., Churazov, E. M., and Krivonos, R. A., 2014, \newblock {\it Mon. Not. R. Astron. Soc.}, {\bf 445}, 1205.

\bibitem{}Riegler, G. R., 1970, \newblock {\it Nature}, {\bf 226}, 1041. 

\bibitem{}Shapiro, S. L., Lightman, A. P., and Eardley, D. M., 1976, \newblock {\it Astrophys. J.}, {\bf 204}, 187.

\bibitem{}Shrader, C. and Titarchuk, L., 1998, \newblock {\it Astrophys. J.}, {\bf 499}, L31.

\bibitem{}Sunyaev, R. A. and Truemper, J., 1979, \newblock {\it Nature}, {\bf 279}, 506.	

\bibitem{}Sunyaev, R. A. and Titarchuk, L. G., 1980, {\it Astron. Astrophys.}, {\bf 86}, 121.

\bibitem{}Tarana A., Bazzano A., Ubertini P., and Zdziarski A. A., 2007, \newblock {\it Astrophys. J.}, {\bf 654}, 494.

\bibitem{}Tarana A., Belloni T., Bazzano A., M\'endez M., and Ubertini P., 2011, \newblock {\it Mon. Not. R. Astron. Soc.}, {\bf 416}, 873.

\bibitem{}Titarchuk, L., 1994, \newblock {\it Astrophys. J.}, {\bf 434}, 570.

\bibitem{}Titarchuk, L., Mastichiadis, A., and Kylafis, N.~D., 1996, \newblock {\it Astron. Astrophys. Supp.}, {\bf 120}, 171.

\bibitem{}Titarchuk, L., Mastichiadis, A., and Kylafis, N.~D., 1997, \newblock {\it Astrophys. J.}, {\bf 487}, 834.

\bibitem{}Titarchuk, L. and Seifina, E., 2009, \newblock {\it Astrophys. J.}, {\bf 706}, 1463. 

\bibitem{}White, N. E., Peacock, A., Hasinger, G., Mason, K. O., Manzo, G., Taylor, B. G., and Branduardi-Raymont, G., 1986, \newblock {\it Mon. Not. R. Astron. Soc.}, {\bf 218}, 129.

\bibitem{}White, N. E., Stella, L., and Parmar, A. N., 1988, \newblock {\it Astrophys. J.}, {\bf 324}, 363.	

\bibitem{}Zdziarski, A. A., Grove, J. E., Poutanen, J., Rao, A. R., and Vadawale, S. V., 2001, \newblock {\it Astrophys. J.}, {\bf 554}, L45.

\end{thebibliography}
\end{document}